# Life Equations for the Senescence Process

Xiaoping Liu*

Davis Heart and Lung Research Institute, Department of Internal Medicine, The Ohio State University College of Medicine, 420 West 12th Avenue, Columbus, OH 43210, USA

Running title: Life Equations

Correspondence to:   Xiaoping Liu
Department of Medicine
The Ohio State University
420 W 12$^{th}$ Ave, 188 TMRF
Columbus, OH 43210, USA
Phone: +1 614-292-1305
Fax: +1 614-292-8778
Email: liu.547@osu.edu

**Keywords:** Mortality kinetics, survival curve, lifespan.

**Abstract**
The Gompertz law of mortality quantitatively describes the mortality rate of humans and almost all multicellular animals. However, its underlying kinetic mechanism is unclear. The Gompertz law cannot explain the effect of temperature on lifespan and the mortality plateau at advanced ages. In this study a reaction kinetics model with a time dependent rate coefficient is proposed to describe the survival and senescence processes. A temperature-dependent mortality function was derived. The new mortality function becomes the Gompertz mortality function with the same relationship of parameters prescribed by the Strehler-Mildvan correlation when age is smaller than a characteristic value $\delta$, and reaches the mortality plateau when age is greater than $\delta$. A closed-form analytical expression for describing the relationship of average lifespan with temperature and other equations are derived from the new mortality function. The derived equations can be used to estimate the limit of average lifespan, predict the maximal longevity, calculate the temperature coefficient of lifespan, and explain the tendency of survival curve. This prediction is consistent with the most recently reported mortality trajectories for single-year birth cohorts. This study suggests that the senescence process results from the imbalance between damaging energy and protecting energy for the critical chemical substance in the body. The rate of senescence of the organism increases while the protecting energy decreases. The mortality plateau is reached when the protecting energy decreases to its minimal levels. The decreasing rate of the protecting energy is temperature-dependent. This study is exploring the connection between biochemical mechanism and demography.

**Introduction**
The human mortality rate can be fit by an exponential formula that was proposed by Benjamin Gompertz in 1825 (1). This formula is known today as the Gompertz "law of mortality" that describes the process of senescence in almost all multicellular animals (2). Searching for the mechanism underlying the mortality law started as early as Gompertz publishing his paper on the law of mortality (3, 4). At the present, there are three main strands of demographic theories, mutation accumulation, stochastic vitality, and optimal life histories, in studying aging process (5). Furthermore, changes in certain substances and inherent energy were also considered to be related to aging process (3, 6-8). Nearly a century ago, it was demonstrated that the ambient temperature might change the lifespan of the aseptic flies of *Drosophila* (9) and that the temperature coefficient for Drosophila's lifespan is close to that of a chemical reaction, indicating that chemical substances play an important role in the aging process. A few years later, Brownlee conjectured that the inherent energy of certain substances in the body is gradually being destroyed throughout life (6). However, in the same paper, he also pointed out "Various formulas in physical chemistry, which might be considered possibly applicable, were tried without success". Up to now, it is unclear whether Brownlee's conjecture is correct or not. Lines of evidence show that the real mortality rate at very advanced ages levels off from the exponential formula of mortality rate (10-16), and the Gompertz law of mortality cannot be used to predict the effect of temperature on lifespan because temperature is not a variable or parameter in the Gompertz function. In this study, the author will explore the following questions: What is the biochemical mechanism underlying Gompertz's law of mortality? Why does a mortality rate plateau exist at very advanced ages? How can the effect of temperature on lifespan be explained? A new mortality rate function will be presented. This new morality function predicts the mortality plateau appearing in the period of age greater than $\delta$, consistent with the most recent human mortality data demonstrated by (17). Analytical expressions for describing the relationship of average lifespan with temperature and some other equations will be derived from this new function.

**Theoretical methods**
Considering that a molecule has an initial concentration $c_0$ and its decomposition rate ($dc/dx$) follows first-order kinetics as below:

$$\frac{dc(x)}{dx} = -kc(x) = -k'(e^{-E_a/RT})c(x) \quad \text{or} \quad \frac{dN(x)}{dx} = -kN(x) = -k'(e^{-E_a/RT})N(x) \quad (1)$$

where $c$ is the concentration of the molecules at time $x$, $k$ is the rate constant, $E_a$ is activation energy for the decomposition reaction, $k'$ is the pre-exponential factor, $R$ is the gas constant, $T$ is the absolute temperature, and $N$ is the number of molecules at time $x$ ($N_0$ is initial number of the molecules). The above two equations can be converted to each other by a factor of the volume V containing these molecules. Assuming that the life process is a chemical process and that a large number of people ($s_0$) with identical biological and chemical properties are born on the same day at time 0, the rate of change in the population ($ds/dx$) is similar to Eq. (1):

$$\frac{ds(x)}{dx} = -ms(x) = -(Ae^{-E_l/RT})s(x) \quad (2)$$

where $m$ is the rate coefficient, $A$ is the pre-exponential factor, and $E_l$ is the activation energy relating to the rate of decrease of the survival function $s$. $E_l$ can be also considered as the living energy that is required for survival. Unlike the normal simple molecules, humans will age and their death rate varies with age, so $m$ is a function of $x$.

*Hypothesis*: The living energy $E_l$, which is proportional to the quantity of a vital molecular unit that linearly decreases with time, is the energy to protect the chemical substance that is critical for life from being impaired by damaging energy (the magnitudes of challenges (7)) from other molecules or from itself. This hypothesis suggests that aging results from the imbalance between damaging energy and protecting energy for the critical chemical substance. The rate of senescence of the organism increases while the protecting energy decreases.

Under this hypothesis, we assume that the initial quantity of this vital molecular unit is $l_0$ and its decrease rate is $b$. At time $x$, the quantity ($L$) of this molecular unit is:

$$L(x)=l_0-bx \qquad (3)$$

The identity of this vital molecular unit is unclear. Telomeres, which protect DNA (the critical chemical substance) from degradation, may be one of candidates. It has been showed that the length of the shortest telomeres is the major determinant of the onset of senescence (18, 19). Since average length of telomeres and the shortest 5% of telomeres are shortened at a rate nearly linear with time (20, 21), we can assume that the length of the shortest telomeres (such as the shortest 5% (21)) also decreases linearly with time. In this case, $L$, $l_0$ and $b$ can be considered as the length of the shortest telomeres at time $x$, the initial length of the shortest telomeres, and the shortening rate of telomeres. Based on the above hypothesis, the living energy $E_l$ is proportional to $L$:

$$E_l(x)=k_0 L(x)= k_0(l_0-bx) \qquad (4)$$

where $k_0$ is the coefficient for converting the length of telomeres into the living energy. Eq. (4) defines the conversion relationship from the vital molecular unit to the living energy. In the real world, a person's death is usually caused by some diseases. The diseases may be prevented by some medicines, physical exercise, and good personal hygiene. If a person becomes ill with a disease, the disease may be medically treated. As a result, these external "forces" can enlarge or reduce the living energy. Therefore, $k_0$ has included the effect of these external "forces".

In Eq. (4), $b$ is a rate constant of telomere shortening, so it can be written in the following form based on Arrhenius equation:

$$b = b_0 e^{-E_t/RT} \qquad (5)$$

where $E_t$ is the activation energy independent of time $x$ in the telomere shortening process.

Eq. (4) can be written as:

$$E_l(x) = k_0(l_0 - bx) = \beta' - \alpha' x = \beta' - (\alpha_0' e^{-E_t/RT})x \qquad (6)$$

where $\beta'=k_0 l_0$ and

$$k_0 b = \alpha' = \alpha_0' e^{-E_t/RT} = k_0 b_0 e^{-E_t/RT} \qquad (7)$$

Since the effective length of the telomere cannot be less than 0, we have:

$$l_0-bx \geq 0 \text{ or } E_l=\beta'-\alpha' x \geq 0 \qquad (8)$$

Substitution of Eq. (6) into Eq. (2) gives:

$$m(x) = A\exp(\frac{\alpha' x - \beta'}{RT}) = A\exp(\alpha x - \beta) = A\exp(\alpha(x-\delta)), \text{ for } x \leq \beta/\alpha=\delta \qquad (9)$$

At $x=\delta$, $E_l$ becomes zero and $m(x)$ becomes the constant $A$. In this situation, $m(x)$ is no longer dependent on time $x$ because the effective telomere length or the living energy has dropped to zero already. So $m(x)$ will remain at the constant $A$ as $x$ continuously increases:

$$m(x) = A, \text{ for } x \geq \delta \qquad (9)'$$

In Eq. (9), $\beta=\beta'/RT$ and

$$\alpha = \alpha'/RT = (\alpha_0'/RT)e^{-E_t/RT} = k_0 b_0 e^{-E_t/RT} \qquad (10)$$

$$\delta = \beta/\alpha = \beta'/\alpha' = l_0/b = l_0/(b_0 e^{-E_t/RT}) = \beta'/(\alpha_0' e^{-E_t/RT}) \quad (11)$$

Based on the definition of mortality rate, $m(x)$ is the mortality rate. Substitution of Eqs. (9) and (9)' into (2) gives:

$$\frac{ds(x)}{dx} = -m(x)s(x) = -Ae^{\alpha(x-\delta)}s(x), \text{ for } x \leq \delta \quad (12)$$

$$\frac{ds(x)}{dx} = -m(x)s(x) = -As(x), \qquad \text{for } x \geq \delta \quad (12)'$$

Solving the differential equations (12) and (12)', we can derive the expression for the normalized survival function $s/s_0$, or $S(x)$

$$S(x) = \frac{s(x)}{s_0} = B\exp(-\frac{A}{\alpha}e^{\alpha(x-\delta)}) = B\exp(-e^{\alpha(x-x_0)}), \text{ for } x \leq \delta; \quad (13)$$

$$S(x) = Be^{-A(x-\delta+1/\alpha)} = Be^{A(\delta-1/\alpha)}e^{-Ax} = B'e^{-Ax}, \qquad \text{for } x \geq \delta \quad (13)'$$

where $B = \exp[(A/\alpha)\exp(-\alpha\delta)]$, $B' = Be^{-A(\delta-1/\alpha)}$, and

$$x_0 = \delta - \ln(A/\alpha)/\alpha, \text{ or } \delta = x_0 + \ln(A/\alpha)/\alpha \quad (14)$$

In many cases, such as for human survival functions, $\alpha\delta \gg 1$, so $(A/\alpha)\exp(-\alpha\delta)$ is close to 0 or the parameter $B$ is close to one. Thus, B can be omitted in the following derivation of equations.

Eq. (13) has the same form as the Gompertz survival function. When $x \geq \delta$, the decay rate of $S(x)$ follows first order kinetics (Eq. (13)'). The decay rate constant is $A$ and the half-life $t_{1/2}$ of $S(x)$ at very advanced ages can be calculated from the following half-life formula for the first-order reaction:

$$t_{1/2} = \frac{\ln(2)}{A} = \frac{0.693}{A} \quad (15)$$

From the above equations, we can see that $\delta$ is an important characteristic value representing the average time that a human body uses up all its living energy. Therefore, $\delta$ can be looked at as the characteristic value for life. Eq. (9) can be rewritten in the form: $m(x) = R\exp(\alpha x)$, where $R = A\exp(-\alpha\delta)$, or

$$\ln(R) = \ln(A) - \alpha\delta \quad (16)$$

Eq. (13) is the same as the SM correlation (linear relationship between $\ln(R)$ and $\alpha$) derived by Strehler and Mildvan (General theory of mortality and aging, 1960) if the symbols $R$, $A$ and $\delta$ in Eq. (16) here are replaced by $R_0$, $K$ and $1/B$ in Eq. (16) in SM's paper (7), respectively. Taking logarithms on both sides of Eq. (9), we have

$$\ln(m(x)) = \ln(A) + \alpha(x-\delta) = (\ln(A) - \alpha\delta) + \alpha x, \text{ for } x \leq \delta \quad (16)'$$

Taking logarithms on both sides of Eq. (9)', we have

$$\ln(m(x)) = \ln(A), \text{ for } x \geq \delta \quad (16)''$$

Eq. (16)' shows that the plot of $\ln(m(x))$ vs. $x$ is a straight line with a slope of $\alpha$ and the intercept of $\ln(A) - \alpha\delta$. At $x = \delta$, all plots of $\ln(m(x))$ vs. $x$ will intersect at the point $(\delta, \ln A)$ regardless of the value of $\alpha$ if both $A$ and $\delta$ are constants. Eq. (16)' is known as compensation law of mortality (22). The mortality plateau will appear when $x \geq \delta$ as described by Eq. (16)''.

## Results

*Finding the life characteristic value $\delta$*

Eq. (13) was used to fit survival curves of 11 countries around the world (23, 24) for the past 100-200 years. Since the data points are not stable at younger ages, we used data points for $\geq 40$ years of ages in the curve-fitting analysis. The parameters $B$, $\alpha$ and $x_0$ were determined from the best fitting lines. These determined parameters were used to further calculate the parameter $\delta$ from Eq. (14). The determined values of $x_0$ and the calculated values of the parameter $\delta$ for the 11 countries (males) are plotted in Fig. 1. These data show that $x_0$ ($x_0$ is proportional to the life expectancy at birth) increased linearly in the past 200 years, but $\delta$ is almost a constant (100.4 ± 2.0 years) in the same period. The average value of $\delta$ between 1820 and 1900 is 101.3±1.4 years. It slightly decreased in the period from 1900 to 1940 and increased from 1960 to the present. The average value of $\delta$ between 1940 and 1960 is 99.0±1.6 years, but it increases to 102.5±1.9 years between 2000 and 2011.

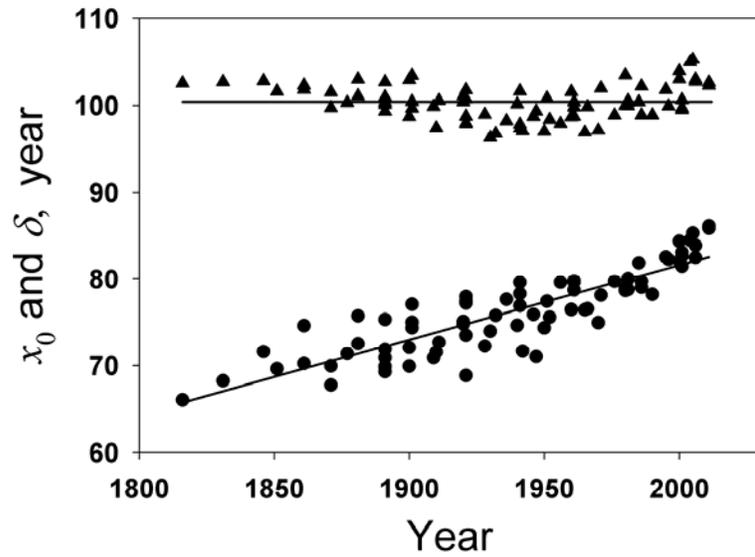

Fig. 1. Determination of parameters $x_0$ and $\delta$. Values of $x_0$ are determined from the survival data of 11 countries (male) (Australia, Canada, Denmark, France, Germany, Japan, Netherlands, Norway, Sweden, Scotland and USA) for the past 100-200 years. The plotted data points show that $x_0$ (●) linearly increases in the past 200 years, but $\delta$ (▲) is almost independent of time.

*Mortality plateau at very advanced ages*

An examination of mortality rate in the United States from 1999 to 2007 shows that the plot of logarithm of mortality rate vs. age ($x$) for the ages greater than 80 years old for each of these 8 years is similar. The logarithm of the mortality rate reaches a plateau at age ~104 (Fig. 2A). The mortality rate plateau was also seen in other animals (11, 25). However, the mortality plateau or deceleration at old age was not clearly seen in a number of studies on mammals (22, 26, 27). One may ask: Is there no mortality plateau at all, or is there a mortality plateau that appears at very advanced ages that was not shown in earlier studies? In a recent paper, Gavrilova and Gavrilov showed that the logarithms of mortality rates for 1895 and 1898 birth cohorts of U.S. women linearly rose with age up to very advanced ages, no mortality plateau appeared for the 1895 cohort, but a mortality plateau appeared for the 1898 cohort (HMD data) after 104 years of age (17). Their data are very helpful for answering the above question. To compare the two mortality trajectories, we read the data points from their paper and re-plotted these data points in Figs. 2B & 2C. The two linear plots intersect at a very advanced age (Fig. 2B). To determine the intersection point, we re-plotted the two sets of mortality data in Fig. 2C, but only the data with ages greater than 80 are shown. The solid lines are the best-fitting lines for the data points (1895 and 1898 birth cohorts) between 80 to 90 years. The two fitting lines intersect at the point (104.2, -0.495) designated by the red star. The dashed line is the mortality plateau predicted from Eq. (16)′′ for the 1898 birth cohort.

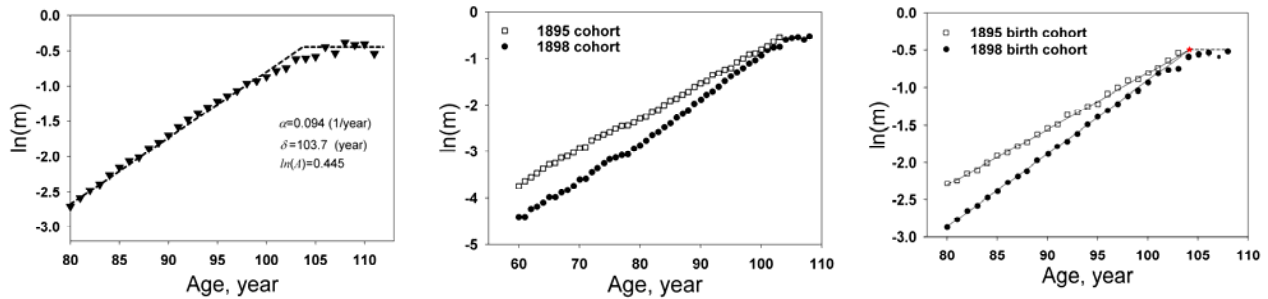

Fig. 2. Mortality rate at advanced ages. A: Mortality trajectory Survival data points (▼) were obtained from 2006 U.S. real death data for the total population. The dashed line was calculated by assuming $\alpha$=0.094 1/year, $\delta$=103.7 year, and ln($A$)=-0.445. B: Mortality trajectories for two single-year birth cohorts of U.S. women (data points were read from the published paper (17)). C: Mortality data sets are the same as those in B, but only the data with ages greater than 80 are shown. The solid lines are the best-fitting lines for the data points (1895 and 1898 birth cohorts) between 80 to 90 years. The two fitting lines intersect at the point (104.2, -0.495) designated by the red star. The dashed line is the mortality plateau predicted from Eq. (16)ʹʹ for the 1898 birth cohort.

*Effect of temperature on survival*
The average lifespan or the life expectancy at birth ($LE_0$) can be calculated from Eqs. (13) and (13)′:

$$LE_0 = \int_0^\infty S(x)dx = \int_0^\infty \exp(-e^{\alpha(x-x_0)})dx + e^{-A(\delta-1/\alpha)} \int_\delta^\infty e^{-Ax}dx - \int_\delta^\infty \exp(-e^{\alpha(x-x_0)})dx \qquad (17)$$

If the mortality plateau occurs at the tail part of a survival curve (such as human survival curve (Fig. 2) or other animals (28)) or the difference of the last two terms in the right side of Eq. (17) is small enough comparing to the first term, then the last two term can be ignored. Under this approximation, $LE_0$ can be approximately expressed by the first term in Eq. (17). Let $y=\alpha(x-x_0)$, $dy=\alpha dx$. Eq. (17) can be rewritten as:

$$LE_0 = \int_0^\infty \exp(-e^{\alpha(x-x_0)})dx = \frac{1}{\alpha}\int_{-\alpha x_0}^\infty \exp(-e^y)dy \qquad (18)$$

For $\alpha x_0 \geq 2$, Eq. (18) can be approximately expressed as:

$$LE_0 = \frac{1}{\alpha}\int_{-\alpha x_0}^\infty \exp(-e^y)dy \approx \frac{1}{\alpha}(\alpha x_0 - 0.577) = x_0 - \frac{0.577}{\alpha} \qquad (19)$$

where the constant 0.577 is the Euler–Mascheroni constant (29). Combination of Eq. (14) with (19) gives:

$$LE_0 = \delta - \frac{\ln(A/\alpha) + 0.577}{\alpha} = \frac{1}{\alpha}(\beta - \ln(\frac{A}{\alpha}) - 0.577) = \frac{Q}{\alpha} = \frac{QRT}{\alpha'} = \frac{QRT}{\alpha_0'}e^{E_t/RT} \qquad (20)$$

where $Q=(\beta-\ln(A/\alpha)-0.577)$. When $T$ varies ±15 °C around 25 °C (298 ± 15 K), the change in $\ln(Q)$ is relatively small. Applying logarithm to both sides of Eq. (20) gives:

$$\ln(\frac{LE_0}{T}) = \ln(g_1) + g_2\frac{1}{T} \qquad (21)$$

where $\ln(g_1)=\ln(QR/\alpha_0')$ is close to a constant independent of $T$, and $g_2=E_t/R$. Eq. (21) is similar to the empirical formula proposed by Shaw et al (8).
Taking the derivative of both sides of Eq. (21) with respect to $T$ and considering that $g_2/T \gg 1$ (based on published experimental data (30, 31)), we can obtain the equation for the temperature coefficient $\theta$ of $LE_0$:

$$\theta = \frac{d(LE_0)}{dT} = \frac{LE_0}{T}(1 - \frac{g_2}{T}) \approx -\frac{g_2}{T^2}LE_0 \qquad (22)$$

The negative sign indicates that $LE_0$ decreases as $T$ increases. Eq. (22) shows that $\theta$ of a species is proportional to the activation energy and $LE_0$ of the species and inversely proportional to $T$ squared.

To test Eq. (21), we examined literature data regarding the effects of temperature on $LE_0$ of poikilotherms such as flies. The results show that $\ln(LE_0/T)$ (▲, Drosophila (30); ●, Calliphora stygia (31)) is linear with $1/T$. From the slope of the best fitting lines (solid lines in Figs. 3A & 3B), we can determine the value of $g_2$, which is close to 10,000 (K). Assuming T=293 K (20 ºC) and $g_2$ is 10,000 K, the calculated temperature coefficient from Eq. (22) is $\theta \approx -0.12 LE_0$. This indicates that the life will increase $\sim 0.12 LE_0$ when temperature decreases by one degree and will double the average lifespan when temperature decreases by eight degrees.

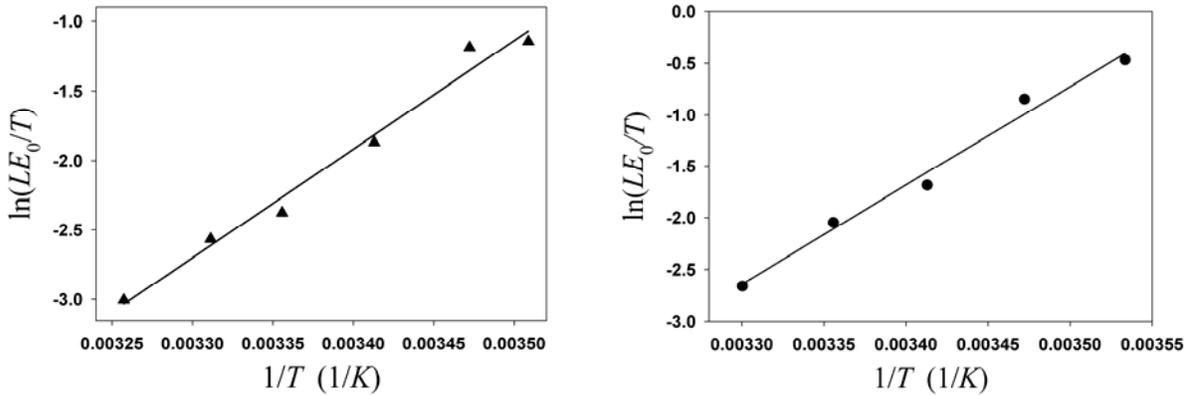

Fig. 3. Effect of ambient temperature on lifespan of flies. A: *Drosophila*; B: *Calliphora stygia*.

## Discussion

Based on the hypothesis on the living energy and the vital molecular unit, such as the average length of the shortest telomeres, the author derived a new mortality rate equation. This new equation has two branches (Eqs. (9) & (9)′) which are divided by the characteristic value for life, $\delta$. At $x \leq \delta$, the equation has the same form as the Gompertz mortality rate equation and leads to the same parameter relationship (Eq. (16)) as the one described by the Strehler-Mildvan correlation. However, the new mortality rate equation contains two new terms. One is the mortality rate in the $x \geq \delta$ branch (Eq. (9)') that predicts the mortality plateau at very advanced ages, and the other is the temperature-dependent parameter (Eqs. (10) and (11)) describing the temperature dependence of the mortality rate function. In addition to Eq. (16), several other equations have been derived from Eq. (9).

The derived Eq. (14) gives the relationship between $\delta$ and other parameters that can be obtained from the survival ($x_0$ and $\alpha$) and mortality ($A$) curves. The values of $x_0$ determined from survival curves of 11 countries and the values of $\delta$ calculated from Eq. (14) are demonstrated in Fig. 1. It can be seen that $x_0$ gradually increases during the period in the past centuries, but $\delta$ is close to a constant (100.4 ± 2.0 years) with a slightly but continuously increase since 1950-1960. If $x_0$ continuously increases and $\delta$ remains at a constant, $\alpha$ will increase based on Eq. (14). In this case, the upper part of the survival curve expressed by Eq. (9) will shift to the right side quicker than the lower tail part. This tendency of change in $x_0$ and $\delta$ is consistent with the tendency (rectangularization (32, 33)) of change in the patterns of human survival curves in the past centuries. An example of Norway survival curves is shown in Fig. 4. This rectangularization is also a result prescribed by Strehler and Mildvan theory of mortality because the theory assumed that $\delta$ (or 1/B) is a constant (7, 34). However, analysis on the survival data of the last century shows that this process continued with a slight tendency to derectangularization of the survival curves in the developed countries after 1950-1960 (34, 35). This derectangularization is deviated from the Strehler and Mildvan theory of mortality. In contrast, the value of $\delta$ is not required as a constant in the new mortality equation presented above, because $\delta$ is dependent on body temperature. It is interesting that this derectangularization process proceeds with the increase in $\delta$ in the same period (after 1950-1960). The new mortality rate equation shows that the age-dependent rate coefficient or mortality rate increases exponentially with age until age reaches the characteristic value of life, $\delta$. As $x \geq \delta$, $\ln(m(x))$ is a constant, $\ln(A)$ (Fig. 2).

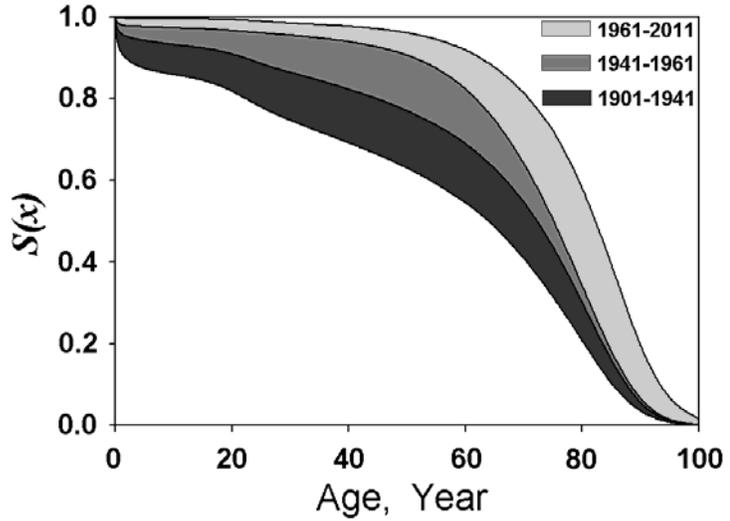

Fig. 4. Period patterns of survival improvement in Norway.

This type of mortality rate can be seen in some organisms such as seed beetle species (10, 36). It was also demonstrated that the logarithm of mortality rate is linear with age up to very advanced ages when mortality data of single-year birth cohorts are used (17). The mortality trajectories for 1895 birth cohort and 1898 birth cohort of U.S. women intersect at 104 years of age (Fig. 2B and Fig. 2C). From Eqs. (16)′ and (16)′′, we know that the value of $x$ at the intersection is equal to $\delta$, and the mortality plateau (the value of y at the intersection) will appear when the age $x$ is greater than $\delta$. It is interesting to note that the mortality plateau does show up for the 1898 birth cohort after 104 years of age, and there is no plateau in the mortality curve for the 1895 birth cohort before 104 years of age. Thus, mortality trajectories shown in Fig. 2B and 2C are an excellent example for the new mortality function (Eqs. (16)′ and (16)′′). To see a mortality plateau for humans, the initial sample size needs to be large enough to ensure enough samples at very advanced ages. Furthermore, human health needs to improve so that more people can survive to very advanced ages. As more and more people can live over 100 years, the mortality plateau will be more easily observed from mortality curves in the future.

The derived Eq. (20) gives the relationship between $LE_0$ and the parameters in survival and mortality functions ($\alpha$, $\delta$ and $A$). Eq. (20) shows that $LE_0$ approaches $\delta$, as $\alpha$ approaches infinity. The average $\delta$ in the past 200 years for the male population is nearly constant (~101 years), indicating that the ratio of $l_0$ to $b$ (the time it takes to use up the living energy) did not significantly extend. Therefore, based on Eq. (20), the $LE_0$ increase in the past 200 years was mainly caused by $\alpha$ relating to those external "forces", such as improvement in health and medical services, which contributed to $k_0$ for enlarging the living energy. This indicates that the maximal $LE_0$ will be limited by the characteristic value of life $\delta$ if biomedical technology in the future can only increase $\alpha$ but are not able to extend $\delta$ ($\delta$ ranges from 102 to 105 years between 2004 and 2011). The mortality will reach its maximal rate at age $\delta$, implying that the mortality will not increase exponentially with age, but will remain in a constant $A$ after age $\delta$. As a result, the mortality rate for human with ages greater than $\delta$ will significantly slower than that predicted by Gompertz mortality law. Despite with this slow mortality rate, if there are 10,000 people at 105 years of age, only one of them may be alive at 120 assuming $A$=0.64. Therefore, the probability of living up to 120 years for a human being is extremely low(37), if $\delta$ is not extended. To significantly extend human maximal lifespan, $\delta$ needs to be elongated and/or the life decay rate constant $A$ needs to decrease. The life decay is likely a characteristic of the human body. It may not be cured by modern medicine at this point (38). However, we hope that this may be changed in the future. Since $\delta$ is the ratio of the average initial length of the shortest telomeres ($l_0$) to the shortening rate of telomeres ($b$), the increase in $\delta$ implies the decrease in $b$ if we assume that $l_0$ does not change in humans. Decrease in $b$ can be reached through either physical methods (such as room temperature) or biochemical methods. It is interesting that the average $\delta$ value has been slowly increasing in the past 50 years from its minimum (1940-1960). From Eq. (5) we know that decrease in $b$ can be caused by the decrease in body temperature, and the decrease in body temperature may be related to the continuous improvement of working and living conditions (such as the broad application of air conditioners in the developed countries). A line of evidence has showed that temperature reduction in both poikilotherms and homeotherms extends lifespan (39-44). To test if the new mortality equation can explain the effect of temperature on lifespan, a simple expression, Eq. (21), for describing the relationship between the average lifespan ($LE_0$) and body temperature was derived from the new mortality rate equation. This expression is similar to the empirical formula proposed by Shaw et al (8). As shown in Fig. 3, the linear relationship of ln($LE_0/T$) with $1/T$ is consistent with published experimental data.

Humans are homeothermic with a core body temperature at 37 ºC. Usually, a living person's body temperature is not uniform (45). Research has shown that when room temperature changes from 35 ºC to 15 ºC, core body temperature can decrease by ~0.7 ºC, and the change in skin temperature is relatively larger (46). Thus, the change in average body temperature should be greater than 0.7 ºC when room temperature varies from 35 ºC to 15 ºC. While humans can normally live at ambient temperature between 15 ºC and 35 ºC without largely life-threatening risk if they have enough water, food, clothes, and suitable medical services, their average body temperature may have 1 ºC difference between the people living in a place with an average temperature of 35 ºC and the people living in another place with an average temperature of 15 ºC. Provided that the $LE_0$ for humans is 75 years and the temperature coefficient for humans is close to poikilotherms (~0.12$LE_0$), the 1 ºC difference in body temperature will result in ~9 years (0.12x75=9) difference in average lifespan under health cold conditions.

**Conclusions**

The author proposed a new mortality rate function. This mortality equation can explain the mortality plateau at advanced age and the effect of temperature on lifespan. Several closed-form analytical expressions have been derived from the new mortality function. Eq. (14) provides a method for determining $\delta$. The determined $\delta$ from 11 developed countries is near a constant in the past centuries, but shows a continuous increase after 1950-1960. This continuous increase in $\delta$ in the past 50-60 years is consistent with the derectangularization of survival curves in the developed countries in the same period. The derived Eqs. (16) and (16)´ have the same forms to the related equations from the Strehler and Mildvan theory of mortality for age $x≤\delta$. The mortality plateau, as described in Eq. (16)´´, appears when $x≥\delta$ . This indicates that the new mortality rate function can not only explain the mortality and survival data for $x≥\delta$ as demonstrated in Fig. 2, but also is consistent with the previous well-known mortality rate equations for $x \leq \delta$. The derived Eq. (21) defines a linear relationship between $\ln(LE_0)$ and $1/T$, which is validated by published experimental data. The derived Eq. (22) is the closed-form expression for temperature coefficient of average lifespan.

**Author Contributions**
X. L. designed this research, derived all equations, directed the process of data search and data collection, analyzed the collected data, prepared all figures, and wrote the whole paper.

**Acknowledgements**
Rachael Huskey is involved in data collection, data analysis and the paper-revising process; and Kerui Liu is involved in the paper-revising process. The author especially thanks Dr. Clay B. Malsh for the discussion with him about the application of mathematical models in health sciences.